\documentstyle[12pt]{article}

\catcode`\@=11
\long\def\@makefntext#1{ 
\protect\noindent \hbox to 3.2pt {\hskip-.9pt
$^{{\ninerm\@thefnmark}}$\hfil}#1\hfill} 

\def\thefootnote{\fnsymbol{footnote}}
 \def\@makefnmark{\hbox to 0pt{$^{\@thefnmark}$\hss}}  
\def\ps@myheadings{\let\@mkboth\@gobbletwo
\def\@oddhead{\hbox{} 
\rightmark\hfil\ninerm\thepage}
\def\@oddfoot{}\def\@evenhead{\ninerm\thepage\hfil 
\leftmark\hbox{}}\def\@evenfoot{}
\def\sectionmark##1{}\def\subsectionmark##1{}}

\begin{document}

\newcommand{\symbolfootnote}{\renewcommand{\thefootnote}
        {\fnsymbol{footnote}}}
\renewcommand{\thefootnote}{\fnsymbol{footnote}}
\newcommand{\alphfootnote}
        {\setcounter{footnote}{0}
--More-- \renewcommand{\thefootnote}{\sevenrm\alph{footnote}}}

\newcounter{sectionc}\newcounter{subsectionc}\newcounter{subsubsectionc}
\renewcommand{\section}[1] {\vspace{0.6cm}\addtocounter{sectionc}{1}
\setcounter{subsectionc}{0}\setcounter{subsubsectionc}{0}\noindent
        {\bf\thesectionc. #1}\par\vspace{0.4cm}}
\renewcommand{\subsection}[1] {\vspace{0.6cm}\addtocounter{subsectionc}{1}
        \setcounter{subsubsectionc}{0}\noindent
        {\it\thesectionc.\thesubsectionc. #1}\par\vspace{0.4cm}}
\renewcommand{\subsubsection}[1]
{\vspace{0.6cm}\addtocounter{subsubsectionc}{1}\noindent
{\rm\thesectionc.\thesubsectionc.\thesubsubsectionc.
        #1}\par\vspace{0.4cm}}
\newcommand{\nonumsection}[1] {\vspace{0.6cm}\noindent{\bf #1}
        \par\vspace{0.4cm}}

\newcommand{\bibit}{\it}
\newcommand{\bibbf}{\bf}
\renewenvironment{thebibliography}[1]
        {\begin{list}{\arabic{enumi}.}
        {\usecounter{enumi}\setlength{\parsep}{0pt}
 \setlength{\leftmargin 0.52cm}{\rightmargin 0pt}
         \setlength{\itemsep}{0pt} \settowidth
        {\labelwidth}{#1.}\sloppy}}{\end{list}}

\topsep=0in\parsep=0in\itemsep=0in
\parindent=1.5pc

\newcounter{itemlistc}
\newcounter{romanlistc}
\newcounter{alphlistc}
\newcounter{arabiclistc}
\newenvironment{itemlist}
        {\setcounter{itemlistc}{0}
         \begin{list}{$\bullet$}
        {\usecounter{itemlistc}
         \setlength{\parsep}{0pt}
         \setlength{\itemsep}{0pt}}}{\end{list}}

\newenvironment{romanlist}
        {\setcounter{romanlistc}{0}
         \begin{list}{$($\roman{romanlistc}$)$}
        {\usecounter{romanlistc}
         \setlength{\parsep}{0pt}
 								\setlength{\itemsep}{0pt}}}{\end{list}}

\newenvironment{alphlist}
        {\setcounter{alphlistc}{0}
         \begin{list}{$($\alph{alphlistc}$)$}
        {\usecounter{alphlistc}
         \setlength{\parsep}{0pt}
         \setlength{\itemsep}{0pt}}}{\end{list}}

\newenvironment{arabiclist}
        {\setcounter{arabiclistc}{0}
         \begin{list}{\arabic{arabiclistc}}
        {\usecounter{arabiclistc}
         \setlength{\parsep}{0pt}
         \setlength{\itemsep}{0pt}}}{\end{list}}

\newcommand{\fcaption}[1]{
        \refstepcounter{figure}
        \setbox\@tempboxa = \hbox{\tenrm Fig.~\thefigure. #1}
        \ifdim \wd\@tempboxa > 6in
           {\begin{center}
        \parbox{6in}{\tenrm\baselineskip=12pt Fig.~\thefigure. #1 }
            \end{center}}
        \else
             {\begin{center}
             {\tenrm Fig.~\thefigure. #1}
              \end{center}}
        \fi}

\newcommand{\tcaption}[1]{
        \refstepcounter{table}
        \setbox\@tempboxa = \hbox{\tenrm Table~\thetable. #1}
        \ifdim \wd\@tempboxa > 6in
           {\begin{center}
        \parbox{6in}{\tenrm\baselineskip=12pt Table~\thetable. #1 }
            \end{center}}
        \else
             {\begin{center}
             {\tenrm Table~\thetable. #1}
              \end{center}}
        \fi}

\def\@citex[#1]#2{\if@filesw\immediate\write\@auxout
        {\string\citation{#2}}\fi
\def\@citea{}\@cite{\@for\@citeb:=#2\do
        {\@citea\def\@citea{,}\@ifundefined
        {b@\@citeb}{{\bf ?}\@warning
        {Citation `\@citeb' on page \thepage \space undefined}}
        {\csname b@\@citeb\endcsname}}}{#1}}

\newif\if@cghi
\def\cite{\@cghitrue\@ifnextchar [{\@tempswatrue
        \@citex}{\@tempswafalse\@citex[]}}
\def\citelow{\@cghifalse\@ifnextchar [{\@tempswatrue
        \@citex}{\@tempswafalse\@citex[]}}
\def\@cite#1#2{{$\null^{#1}$\if@tempswa\typeout
        {IJCGA warning: optional citation argument
        ignored: `#2'} \fi}}
\newcommand{\citeup}{\cite}

\def\fnm#1{$^{\mbox{\scriptsize #1}}$}
\def\fnt#1#2{\footnotetext{\kern-.3em
        {$^{\mbox{\sevenrm #1}}$}{#2}}}

\def\etaten{\eta_{10}}
\def\he{${\rm {}^4He\:\:}$}
\def\be{\begin{equation}}
\def\ee{\end{equation}}
\def\bc{\begin{center}}
\def\ec{\end{center}}
\def\bt{\begin{tabular}}
\def\et{\end{tabular}}
\def\mc{\multicolumn}
\def\beq{\begin{eqnarray}}
\def\eeq{\end{eqnarray}}
\def\beqd{\begin{eqnarray*}}
\def\eeqd{\end{eqnarray*}}
\def\nin{\noindent}
\def\lra{$\leftrightarrow$ }
\def\eset{$\not\!\!\:0$ }
\def\bull{$\bullet$}
\def\reaceight{$^3$He$(\alpha , \gamma )^7$Be}
\def\reacnine{$^3$H$(\alpha , \gamma )^7$Li}
\def\dhe{{\em D + $^3$He }}
\def\dheh{{\em (D + $^3$He)/H }}
\def\omb{\Omega_B}

\font\twelvebf=cmbx10 scaled\magstep 1
\font\twelverm=cmr10 scaled\magstep 1
\font\twelveit=cmti10 scaled\magstep 1
\font\elevenbfit=cmbxti10 scaled\magstephalf
\font\elevenbf=cmbx10 scaled\magstephalf
\font\elevenrm=cmr10 scaled\magstephalf
\font\elevenit=cmti10 scaled\magstephalf
\font\bfit=cmbxti10
\font\tenbf=cmbx10
\font\tenrm=cmr10
\font\tenit=cmti10
\font\ninebf=cmbx9
\font\ninerm=cmr9
\font\nineit=cmti9
\font\eightbf=cmbx8
\font\eightrm=cmr8
\font\eightit=cmti8


\rightline{{\bf CWRU-P10-94}}
\centerline{\tenbf BIG BANG NUCLEOSYNTHESIS AND DARK MATTER:}
 \baselineskip=18pt
\centerline{\tenbf BARYONS AND NEUTRINOS
\footnote{Invited lecture to appear in Proceedings of International
Conference on Critique of the Sources of Dark Matter in the
Universe.}
}
\baselineskip=16pt
\vspace{0.3cm}
\centerline{\tenrm LAWRENCE M. KRAUSS}
\baselineskip=13pt
\centerline{\tenit Department of Physics,
 Case Western Reserve University, 10900
Euclid Ave.}
\baselineskip=12pt
\centerline{\tenit Cleveland OH 44106-7079 U.S.A.}
\vspace{0.5cm}
\centerline{\tenrm ABSTRACT}
\vglue 0.3cm
{\rightskip=3pc
 \leftskip=3pc
 \tenrm\baselineskip=12pt\noindent
I present a review of Big Bang Nucleosynthesis,
concentrating on the statistical analysis of theoretical
uncertainties, and on systematic errors in observed abundances. Both
have important implications for constraints on the amount of baryonic
dark matter and the number of light neutrino species in nature.
Without allowing for systematic uncertainties in abundances, we find
that homogenous BBN would lead to the constraint $\Omega_B \le .07$
and $N_{\nu} \le 3.07$.  Even allowing for maximal systematic uncertainties
in
$^4$He, and $^7$Li, we find  $\Omega_B \le .163$.   For intermediate
ranges we provide new analytic fits for the upper limits to $N_{\nu}$ and
$\Omega_B$ as a function of the helium mass fraction, $Y_p$.
\vglue 0.1cm}

\vfil
\twelverm   
\baselineskip=14pt
\section{Overview}
\vspace*{-0.35cm}
In this review I describe recent work on Big Bang Nucleosynthesis (BBN)
involving an updated Monte Carlo Big Bang Code and new statistical
tools which allow a refined assessment of BBN constraints.  We find that
with these tools the constraints derivable on the baryon abundance
today, and also the effective number of neutrino species can be quite
severe, ruling out for example the possibility of sufficient baryonic
dark matter to make galactic halos, and also ruling out a great deal of
new physics involving new particles beyond the standard model.  The
way out of this connundrum is to allow for systematic uncertainties
in the inferred light element abundances.  One such possibility is
that a recent D observation which indicates substantially more primordial D
than previous analyses suggested is correct.  In this case, the BBN upper
limit on $\Omega_B$ would be substantially reduced, allowing essentially
no baryonic dark matter, but the limit on $N_{\nu}$ would be relaxed.
Another perhaps more likely possibility is that the previously assumed
upper limit on $^4$He should be relaxed, in which case both $\Omega_B$
and $N_{\nu}$ upper limits would be relaxed.  Much of the discussion
provided here is taken from
\cite{kk1,kk2,kk3}, where the reader can turn to for further details.

\section{Code Updates, Elemental Correlations, and Statistical
Uncertainties}
\vspace*{-0.35cm}

 The
remarkable agreement of the predicted
primordial light element abundances and those inferred from present
observations yields some of the strongest evidence in favor of a
homogeneous FRW Big Bang cosmology.  Because of this, significant efforts have
taken place over 20 years to refine BBN predictions, and the
related observational constraints.  Several factors have
contributed to the maturing of this field, including the
incorporation of elements beyond \he in comparison between theory and
observation\cite{who}, and more recently: an
updated BBN code \cite{kawano}, a more accurate measured neutron half
life\cite{helium}, new estimates of the actual primordial \he, $D
+^3He$, and $^7Li$ abundances \cite{walker,con}, and finally the
determination of BBN uncertainties via Monte Carlo analysis
\cite{kraussrom}.  All of these, when combined together\cite{smithetal},
yield a consistent and strongly constrained picture of homogeneous BBN.

We recently returned to re-analyze BBN constraints initially
motivated by three factors:  new measurements of
several BBN reactions, the development of an improved BBN code,
and finally the realization that a correct statistical determination of BBN
predictions should include correlations between the different elemental
abundances. Each serves to further restrict the
allowed range of the relevant cosmological observables $\Omega_{B}$
and $N_{\nu}$. Of course, statistically determined uncertainties are
not the major factor limiting our ability to use BBN to constrain
fundamental parameters.  As we shall see, systematic uncertainties in
the inferred light element abundances are generally much larger, and
must be properly accounted for if we are to conservatively compare
predictions with observations.  In this first section I outline the
details of our effort to properly update and account for BBN statistical
uncertainties, and leave the discussion of systematics to a later
chapter.

\subsection{New BBN Reaction Rates:}
By far the most accurately
measured BBN input parameter is the neutron half-life, which governs
the strength of the weak interaction which interconverts neutrons and
protons.  Since this effectively
determines the abundance of free neutrons at the onset of BBN, it is crucial
in determining the remnant abundance of \he.  With the advent of neutron
trapping, the uncertainty in the neutron half-life quickly
dropped to less than $0.5 \%$ by 1990.  Nevertheless, it is the uncertainty
in this parameter that governs the uncertainty in the predicted \he
abundance.  The world average
for the neutron half-life is now\footnote{In fact as of this writing, the
value of $ \tau_N$ has just been updated\cite{helium}
 to be $887 \pm 2 sec$. Unless otherwise stated,
 this value is now used in
the tables and formulae presented here, which thus update
values found in some of our earlier work\cite{kk1,kk2}.
  The updated tables can in any case also
be found in \cite{kk3}.}
 $\tau_N = 889
\pm 2.1 sec$ \cite{helium}, which has an uncertainty almost twice as
small as that used in previous published BBN analyses
\cite{walker,kraussrom,smithetal}.  We utilized the updated value in our
analysis.  Other updated rates include one based a new measurement of
 $^7Be + p \rightarrow  \gamma + ^8B $ which is about\cite{gai}
$20
\%$ smaller than previous estimates.  One might expect that at high values
of
$\etaten$ (defined by $\Omega_{B} =.0036 h^{-2} (T/2.726)^3
\etaten \times 10^{10}$, where $T$ is the microwave background temperature
today, and $h$ defines the Hubble parameter $ H= 100h$ km/(Mpc sec))
lowering this rate would result in more remnant $^7Li$.  However reducing
the rate by
$20 \%$ in our code alters remnant
$^7Li$ by less than one part in $10^5$!  Other than these two new rates we
used the reaction rates and uncertainties from
\cite{smithetal}.
\vskip 0.2in

\subsection{New BBN Monte Carlo:}
Because
of the new importance of small corrections to the \he abundance when
comparing BBN predictions and observations, increased attention has been paid
recently to effects which may alter this abundance at the $1 \%$ level or
less.  In the BBN code several such effects were incorporated, resulting in
an $\etaten$ -independent correction of $+.0006$ to the lowest order value of
$Y_p$ (the \he mass fraction).  This is a change of $+.0031$ compared to the
value used in previous published
analyses\cite{walker,kraussrom}.

In the present code, more than half of the new correction
is due to finer integration of the nuclear
abundances. Making the time-step in the
code short enough that different Runge-Kutta drivers result in the same
number for the \he abundance produces a nearly $\etaten$ independent
change in $Y_p$ of +.0017 \cite{kernan}. Residual numerical uncertainties
 are small
compared to the uncertainty in $Y_p$ resulting from that in $\tau_N$
 \cite{kernan,KSW}.  The other major change is the
inclusion of ${M_N^{-1}}$ effects\cite{seck:accbbn}. Seckel
showed that the effects on the weak rates due to nucleon recoil, weak
magnetism, thermal motion of the nucleon target and time dilation of
the neutron lifetime combine to increase $Y_p$ by $\sim$ .0012.
Also included in the correction
is a small increase of $.0002$ in $Y_p$ from momentum dependent neutrino
decoupling \cite{turndod,ftd:accbbn}.

Finally, we utilized a Monte Carlo procedure in order to incorporate
existing uncertainties and determine confidence limits on parameters.  Such a
procedure was first carried out\cite{kraussrom} with
BBN reaction rates chosen from a (temperature-independent) distribution
based on then existing experimental uncertainties. This procedure was
further refined\cite{smithetal} by updating
experimental uncertainties and using temperature dependent
uncertainties.  Here we utilized the nuclear
reaction rate uncertainties in \cite{smithetal} (including the
temperature dependent uncertainties for
 \reaceight $\:$ and \reacnine) except for the reactions we updated.
Each reaction
rate was determined using a Gaussian distributed random variable
centered on unity, with a $1-\sigma$ width based on that
quoted in \cite{smithetal}. For the
rates without temperature-
dependent uncertainties this number was used as a multiplier
throughout the
integration. For the two
rates with temperature-dependent uncertainties the original uniformly
distributed random number was saved and mapped into a new gaussian
distribution with
the appropriate width
at each time step.

The results of the updated BBN Monte Carlo analysis are displayed in figure
1, where the symmetric $95 \%$ confidence level predictions for each
elemental abundance are plotted. Also shown are previously claimed
observational upper limits for each of the light elements
\cite{walker,con,smithetal}.  {\it In the first instance we shall utilize
these limits in order to assess how BBN constraints have evolved based on
our re-analysis, and after this we shall consider the more realistic case
where systematic uncertainties are also accounted for.}  Figure 1 also
allows one to  assess the significance of the corrections we used, in
relation to the width of the 95\% C.L. band for
$Y_p$, which turns out to be
$\sim$ .002. The total change in $Y_p$ of $\approx +.003$ from previous BBN
analyses conspires with the reduced uncertainty in
the neutron lifetime, which narrows the uncertainty in
$Y_p$ and feeds into the uncertainties in the other light elements,
 to reduce the range where the
predicted BBN abundances are consistent with the inferred
primordial abundances.

\vglue 3.5in 
\includegraphics{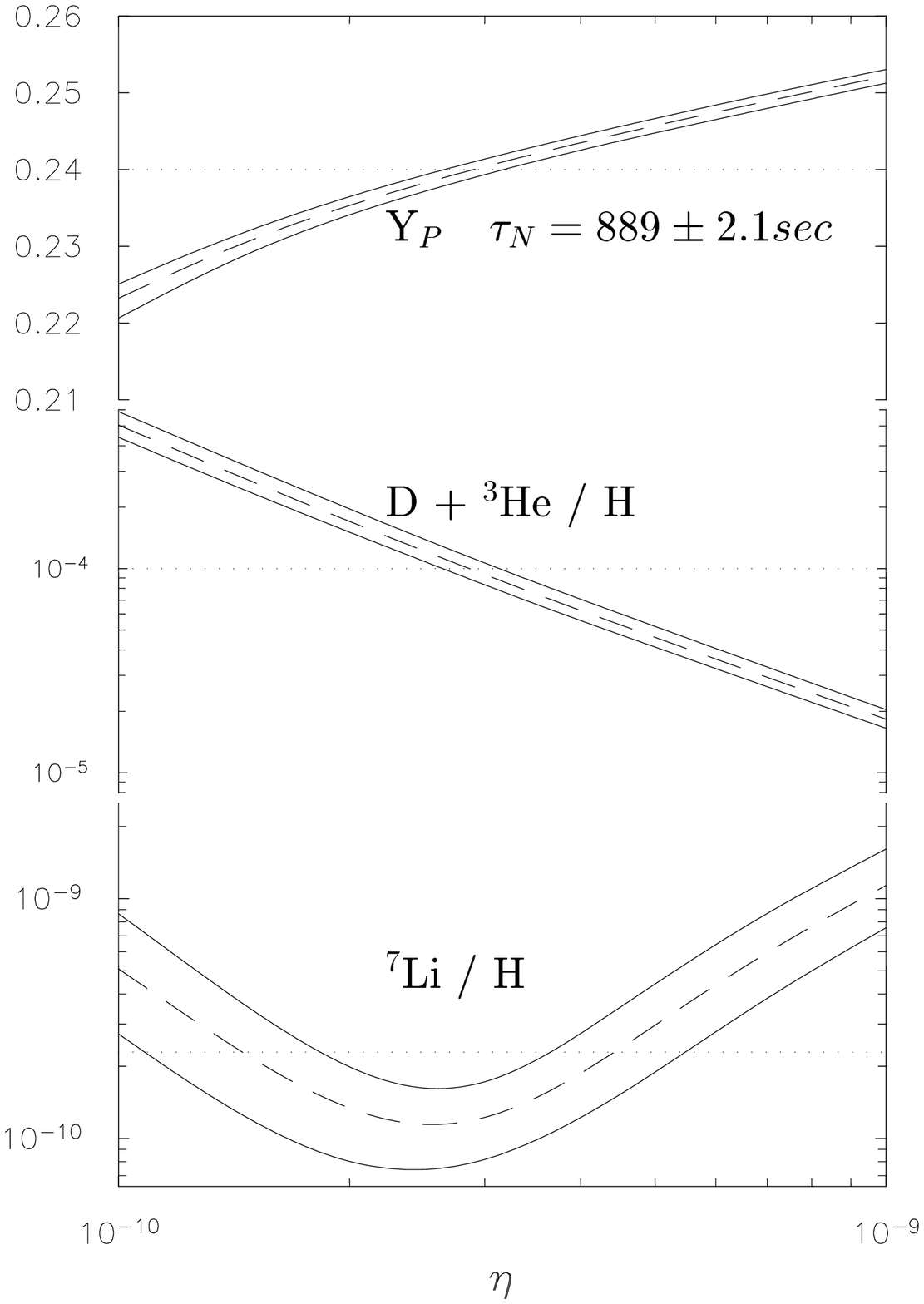}

        Figure 1: BBN Monte Carlo predictions as a function of
$\etaten$.  Shown are symmetric  $95 \%$ confidence limits on each elemental
abundance.  Also shown are claimed upper limits inferred from observation.

\subsection{Statistical Correlations Between Predicted Abundances:}
While the introduction of a Monte Carlo procedure was an important step,
the
determination of limits on the allowed range of BBN parameters
$\Omega_{B}$ and
$N_{\nu}$ based on comparison of symmetric $95 \%$ confidence limits
for single elemental abundances with observations, as has become the
standard procedure, overestimates the allowed range.  This is because the
BBN reaction network ties together all reactions, so that the predicted
elemental abundances are not statistically independent. In
addition, the use of symmetric confidence limits is too conservative.
Addressing these factors is a central feature of our work.

Figure 2 displays the locus of predicted values for the fractions $Y_p$
and D +$^3$He/H for 1000 BBN models generated from the distributions
described above for $\etaten =2.71$ (figure a) and $\etaten =3.08$ (figure
b).  Also shown is the $\chi^2 =4$ joint confidence level contour derived
from this distribution, in a Gaussian approximation, calculating variances
and covariances in the standard manner.  The horizontal and vertical
tangents to this contour correspond to the individual symmetric $\pm
2\sigma$ limits on Gaussianly distributed $ x$
and $y$ variables. As can be seen, the distribution is close to Gaussian,
but has deviations.  Nevertheless, this approximation is useful to
quantify the magnitude of correlations and variances. As is evident from the
figure, and as is also well known on the basis of analytical arguments, there
is a strong anti-correlation between $Y_p$ and the remnant D +$^3$He
abundance (the normalized covariance ranges from -0.7 to -0.4 in the
$\etaten$ range of interest).   Thus, those models where
\he is lower than the mean, and which therefore may be allowed by an upper
bound of $24\% $ on $Y_p$, will also generally produce a larger remnant
D+$^3$He/H abundance, which can be in conflict with the bound on this
combination of
$10^{-4}$ \cite{Walker}. This will have the effect of reducing the
 parameter space
which is consistent with both limits.

\vglue 3in 
\includegraphics{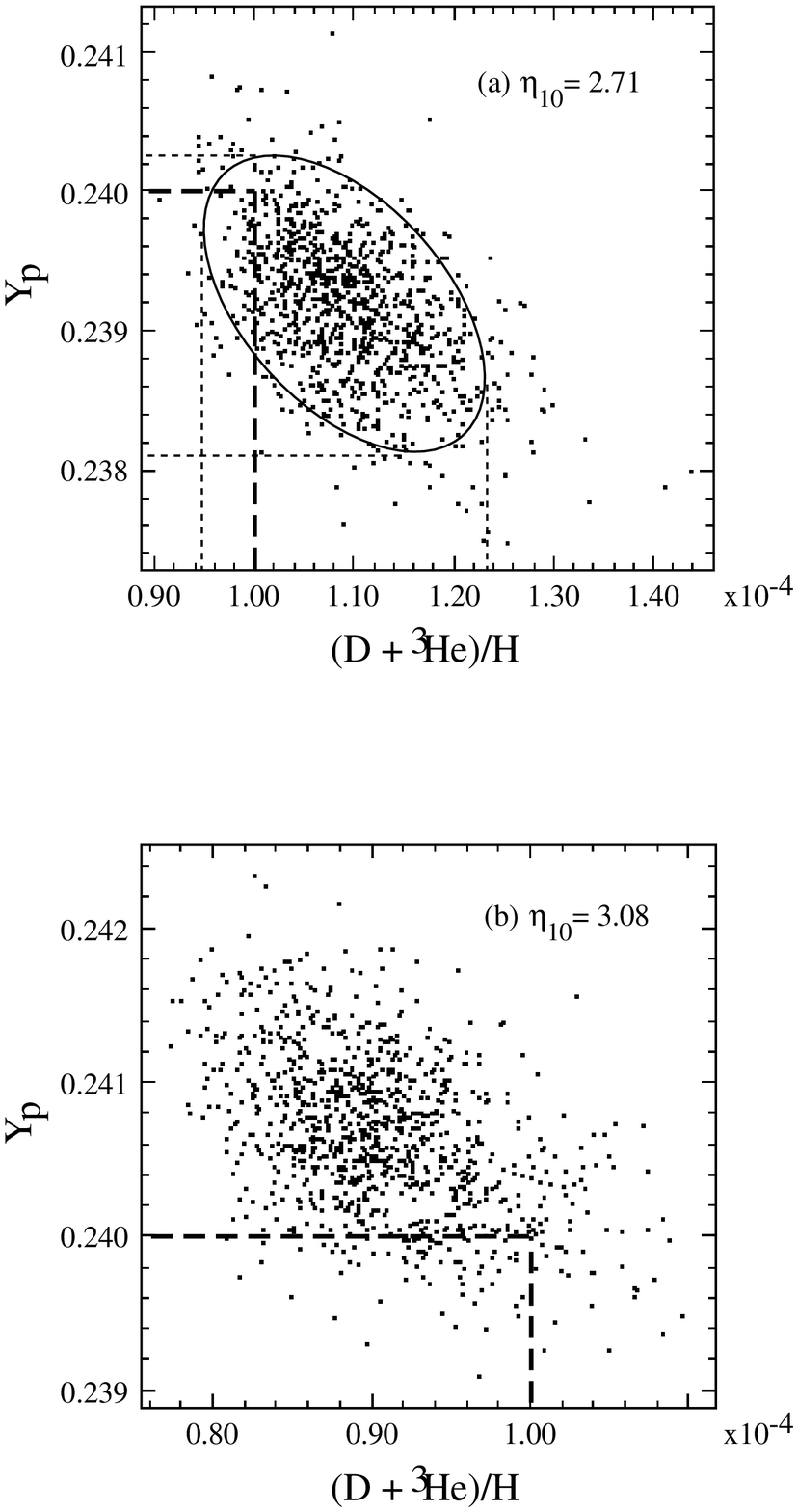}

        Figure 2: Monte Carlo BBN predictions for $Y_p$ vs $D + ^3He$ and
allowed range (using 1992 neutron half life value)
 for (a) $\eta_{10} =2.71$, and (b) $\eta_{10} =3.08$.
In (a) a Gaussian contour with $\pm 2 \sigma$ limits on each individual
variable is also shown.

Because our Monte Carlo
generates the actual distribution of abundances, Gaussian or not, we
 determined
a $95 \% $ confidence limit on the allowed range of $\etaten(N_{\nu})$
by requiring that at least 50 models
out of 1000 lie within the joint range bounded by both the \he and D +$^3$He
upper limits, as shown in figure 2. This
is to be compared with the procedure which one would follow without
considering joint probability distributions.  In this case, one would
simply check whether 50 models lie {\it either} to the left of the
D +$^3$He constraint for low $ \etaten$ (figure a), or below the \he
constraint for high $\etaten$ (figure b).  This is clearly a looser
constraint than that obtained using the joint distribution.  Finally,
the procedure which has been used to-date, which is to check whether the
symmetric $2 \sigma$ confidence limit (i.e. when 25 models exceed either
bound) for a single elemental abundance crosses into the allowed region gives
even a looser constraint, as can be seen in figure 2a.

In table 1 and figure 3 we display these results.  Here we show
the
$95 \% $ confidence limits on $\etaten$, both as we have defined them above
and also using
the looser procedures which ignore correlations.  Accounting for the
correlations in the non-symmetric $95 \%$ confidence limit tightens
constraints by reducing the overall number of acceptable models. This effect
is most significant when the peak probability
(as a function of $\etaten$ in
fig 3) is such that the $95 \%$ confidence
line intersects the distribution
near the peak rather than the tail.  As a result the constraints tighten
dramatically as the number of effective light neutrino species,
$N_{\nu}$ is increased. Assuming the upper limits on
\he and D +$^3$He quoted
above, greater than
$3.04$ effective light neutrino types are ruled out only once
 correlations are taken
into account.  Without including correlations the upper limit would be
$3.15$ neutrino species.

\bigskip
\bc
{Table 1: Correlations \& $\eta_{10}$ Limits (for 1992 neutron half life
 value)}
\vskip .25 truein
\begin{tabular}{|l||c|c|c|c|}  \hline
{95\% C.L.}&\multicolumn{4}{|c|}{$N_{\nu}$} \\ \cline{2-5}
 $\eta_{10}$ range & 3.0 & 3.025 & 3.04 & 3.05 \\ \hline
w/ corr. & 2.69 \lra 3.12 & 2.75 \lra 2.98 & 2.83 \lra 2.89 & \eset \\
\cline{1-1}
w/out corr. & 2.65 \lra 3.14 & 2.65 \lra 3.04 & 2.69 \lra 2.99  & 2.69 \lra
2.95 \\ \cline{1-1}
sym. w/out corr. & 2.62 \lra 3.17 & 2.63 \lra 3.10 & 2.65 \lra 3.03 & 2.66 \lra
3.00 \\ \hline
\end{tabular}
\ec
\bigskip

\vglue 2.75in 
\includegraphics{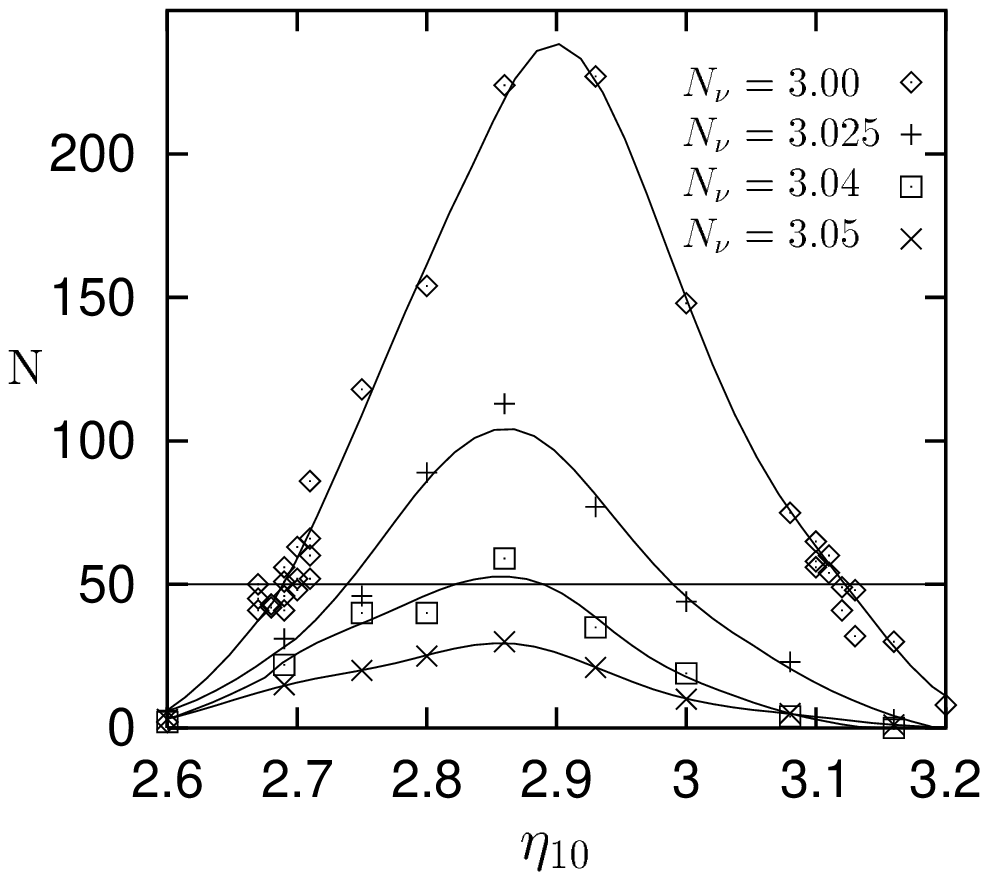}

        Figure 3: Number of models (out of 1000 total models)
which satisfy constraints
$Y_p \le 24 \%$ and $ D + ^3He/H \le 10^{-4}$ as a function of $ \etaten$, for
$3.0, 3.025, 3.04, 3.05$ effective light neutrino species, based on
1992 neutron half life value.  Curves are
smoothed splines fit to the data.

We also determined an
upper limit on
$\etaten$ using just $^7Li$. Requiring
$^7Li/H
\le 2.3 \times 10^{-10}$ \cite{con,smithetal} yields a limit $\etaten
\le 5.27$ .  This is weaker than the \he limit, and there remains
debate about the actual observational upper limit on
primordial $^7Li$.

\subsection{ Implications and Caveats:}

The above
constraints on $\etaten$ and $N_{\nu}$, taken at
face value, assuming a $Y_p$ upper bound of $24\%$ and an D+$^3$He/H
upper limit of $10^{-4}$, would have significant implications
for cosmology,  dark matter and particle physics.  The limit on $\etaten$
corresponds to the limit $0.015 \le \Omega_{B} \le 0.070$\footnote{Using
the
new neutron half life value would raise the upper limit by about $5\%$}.
  (Assuming
 $ 0.4 \le h \le 0.8$, as is required by direct measurements and
limits on the age
of the universe.) Thus, { \it if the previously quoted observational upper
limits on the $Y_p$ and D+$^3$He/H are used directly},  homogeneous BBN
would imply:
\vskip 0.05 in

\noindent (a)  The upper limit on $\Omega_{B}$ would be marginally
incompatible with even the value of $\approx$ 0.1 inferred from  rotation
curves of individual galaxies, further suggesting the need for
non-baryonic dark matter in these systems.
\vskip 0.05 in

\noindent (b)  The bound on the number of effective light degrees of
freedom during nucleosynthesis is {\it very severe}, corresponding to
less than 0.04 extra light neutrinos\footnote{this becomes $\approx .07$
neutrinos using the new neutron lifetime (see section 4)}.
   This is perhaps the most
worrisome bound of all because it rules out {\it any} Dirac mass for a
neutrino without some significant extension of the standard model.  This
is because even a light ``sterile" right handed component whose
interactions freeze out about 300 GeV will still contribute in excess of
$0.047$ extra neutrinos during BBN \cite{Kolbturn} without extra particles
whose annihilations can further suppress its abundance compared to its
original thermal abundance.  It can have further implications
for a variety of kinds of hot or cold dark matter. For example, new light
scalars would be ruled out unless they decouple above the electroweak
scale, as would be any significant population of supersymmetric particles
during BBN. Even allowing  $0.047$ extra light neutrinos, the upper limit
on a Dirac mass would be reduced to
$ \approx 5$  keV
\cite{fullermal,krauss}.
 A $\nu_{\tau}$ mass greater than 0.5 MeV with lifetime exceeding 1 sec.
would also be ruled out due to its effect on the
expansion rate during BBN\cite{turner,kkkssw}.  Also,
neutrino interactions
 induced by  extended technicolor at scales less than $O(100)$ TeV are
ruled out \cite{kraternap}, and sterile
right handed neutrinos \cite{dodelmal} would be ruled out as warm dark matter
as the lower limit on their mass would now be
$O(1 keV)$.

\noindent (c) The primordial $^4$He mass fraction would have to be
greater than 23.8 $\%$ for consistency with D+$^3$He constraints.  This
is very close to the upper limit of 24 $\%$ used in previous analyses.

These constraints are so stringent that they cry out for a consideration of
uncertainties in the light element abundance estimates.  Indeed, as we
shall next discuss, in spite of the considerable effort devised above to
accomodate statistical uncertainties in the predictions, by far the
largest and most significant uncertainties in the comparison of BBN
predictions and observations come from the latter.  Moreover, these
uncertainties are systematic and not statistical.  Accomodating them in
a BBN analysis will be the subject of the rest of this review.

\section{New D Observations?}

As we have just indicated, the weakest link in a BBN analysis involves the
assumed light element abundances.  Estimates of \he, for example
\cite{olive91,pagel92,pagel93} are mostly
indirect, and subject to large systematic uncertainties, which
may also be important for the other abundance estimates. As a result, our
refined BBN analysis described above suggests the need
for revision of the light element abundance estimates inferred from
observation at least as much as it argues for or against new
nonstandard physics.

One of the most worrisome aspects of the present constraints is the fact
that D+$^3$He provides a lower limit on $Y_p$ which is uncomfortably
close to the previously claimed upper limit.  There are obviously two
ways out of this dilemma: either the D+$^3$He limit is increased, which
would allow smaller values of $Y_p$ to be consistent, or the
observational upper limit on
$Y_p$ is increased.  Both possibilities have recently been discussed.  In
this section we treat the former, and in the final section we discuss
the latter.

As far as the possibility of raising the D+$^3$He abundance estimate is
concerned, a new claimed observation\cite{SCHR}, of
deuterium in a primordial gas cloud, at a level
(D+$^3$He)/H
$=1.9-2.5
\times 10^{-4}$ is particularly exciting.  It has long been
argued that any present measurement of {\em D} provides a lower limit
on its primordial abundance because {\em D} is so fragile that it is
easily destroyed in stars.  Also, because the predicted BBN
 abundance falls monotonically with increasing
baryon density, a lower limit on deuterium thus places a reliable
upper limit on the baryon density of the universe.  Previously quoted
abundance estimates of $10^{-5}$ led to a firm upper bound on $
\eta_{10} <8$
which clearly established that baryons could not close the universe.
The Songaila {\it et al.} observation, an order of magnitude larger, is also
a factor of two greater than the previous upper limit on the combination
D+$^3$He.  As a result, this would allow smaller values of $\etaten$, which
would in turn allow a smaller value of $Y_p$, although the upper limit on
$\Omega_B$ one might derive would be much more severe.  In addition, as we
shall show, this result, if upheld, would change the way we combine
elemental abundance limits to get constraints on cosmology and particle
physics.

The system explored by Songaila {\it et al.} in principle provides a
direct probe of unprocessed {\em D}.  Nevertheless, their
measurement is not yet definitive, and could be subject to systematic
errors \cite{SCHR}.  For example, an intervening gas
cloud moving relative to the first cloud at a small radial velocity
could produce {\em H} absorption lines which are shifted, and could mimic
{\em D} absorption lines.  Moreover, while one might expect the {\em
D} abundance in the primordial clouds would exceed that in the galaxy,
it is not clear how the large value obtained can be reconciled with
previous galactic estimates\cite{dear,olive90} .

Caveats notwithstanding, because of its potential importance for
altering BBN constraints it is worthwhile to examine just how these
constraints would change based on the present measurement.
While the authors of
\cite{SCHR} pointed out that at worst their observation provides
a conservative upper bound on D, we have argued that it is appropriate,
from the point of view of BBN to use the lower limit of $1.9 \times
10^{-4}$ on the D fraction as a {\it lower limit} on the primordial
deuterium abundance.  If their result is correct, the
primordial D abundance must exceed this value, since it can only be
destroyed by processes since BBN.  If their result is in error, then it
is not clear that the old upper limit on \dhe should be abandoned just
because
it is less conservative.

In figure 4 we display the predicted ranges for D, \he, and $^7$Li for
the range $ 1< \etaten <2$, along with the new D observational lower
limit, and the previously claimed upper limits on \he and $^7Li$, assuming
3 light neutrino species.  For each value of $\etaten$ all 1000 model
predictions are shown, along with the median predictions, the
one-side
$2
\sigma$ upper (lower) limit (dashed line) for D ($^7Li$), and the
symmetric $2
\sigma$ range for each element (triangle).  The one-side limits occur
when less than 50/1000 models fell below(above) the limits,
while the symmetric limits encompass the central 950/1000 predictions.
The D
and
$^7$Li limits shown would imply the allowed range $ 1.13 <\etaten < 1.87$,
which would lead to a strong constraint on $\Omega_B$. (If
one were to put an upper limit on primordial D of 2.5 $\times 10^{-4}$,
derived from the new D observation alone, the lower limit on $\Omega_B$
would increase  by 15 $\%$ from the more conservative limit quoted above
using $^7$Li.)

\vglue 3in 
\includegraphics{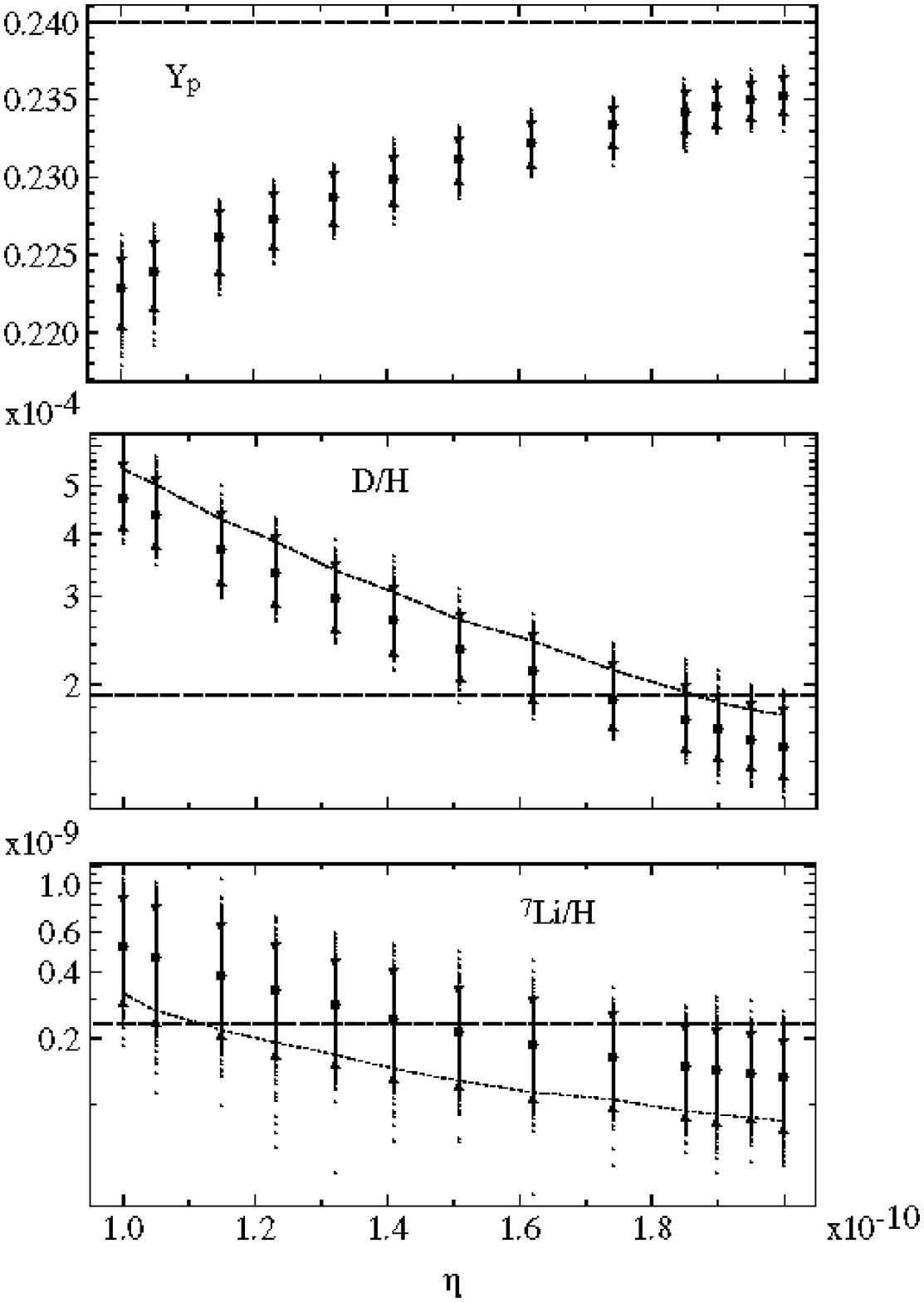}

        Figure 4: Predicted ranges for D, \he, and $^7$Li for
the range $ 1< \etaten <2$, along with the possible new D observational
lower limit.

By considering \he vs D abundances for the upper
limit $\etaten = 1.87$, assuming three species of light neutrinos we
would find that for the region above the Songaila D lower limit,
the maximal value of
\he is near 23.5$\%$.  This is a complete reversal of the ``standard" BBN
limits, which put a lower bound on \he.  What is perhaps more interesting
is that this new upper bound would be far more consistent with the best fit
estimates of
$ 23 \pm 1 \% (2 \sigma)$ which are often applied to \he.

The role played by
varying the ``number" of neutrinos would be quite different than it was
when one combined an upper limit on D+$^3He$ with an upper limit on $^4$He,
aside from a relaxed upper bound on N$_{\nu}$ due to
the lower predicted
\he fraction in this range of
$\etaten$.  Before,
raising the number of neutrinos tightened BBN constraints, but now
it would actually relax them. In order to give the most conservative
limits on
$\etaten$ one must then allow the effective number of neutrinos to vary
from 3, to account for possible new particles contributing to the radiation
gas during BBN at the fraction of a neutrino level.

The point is that in the range of $\etaten$ of interest if the new D
result is confirmed, increasing
$N_{\nu}$ monotonically increases all elemental abundances.
At $\etaten=1.8$ an
increase of $N_{\nu}$ of 1 produces an increase of 5$\%$ in Y$_p$, 15$\%$ in
D/H and 23$\%$ in $^7$Li.
This implies that an increase in the effective number of relativistic
species will increase the upper limit on $\Omega_B$ because the predicted
D abundance will increase.  However, at some point the upper bound on
$\Omega_B$ from \he (or $^7$Li) will eclipse that from D.
By varying the number of
neutrinos, we find the maximum allowed value of $\etaten$ is 1.91, obtained for
3.4 effective neutrino species.

Increasing the number of neutrino species beyond 3.9 would result in
violation of the observational limits for any range of
$\etaten$.  Clearly, any further relaxation
of the $^4$He and $^7$Li limits would then further weaken the bound,
allowing 4.0 neutrinos.

This new D measurement could be quite exciting for
cosmology.  If confirmed, it would change the way we use primordial element
abundances to get cosmological and particle physics constraints, as we have
shown in the new confidence ranges we have derived.  In some sense, the new
cosmological constraints would be satisfying.  The allowed range of
$\Omega_B$ would overlap with the amount of visible matter in the Universe.
What you see may be what you get, a result which has some attraction.
In addition, the  \he fraction can be much closer to what some people
have been claiming, and the constraint on plausible extra particles in the
radiation gas during BBN is relaxed.

Preliminary reports of other similar measurements along different lines
of sight suggest the Songaila {\it et al} value may be too high, in which
case the scenario discussed in this section, however attractive, would
have to be abandoned.  The next place to turn for a relaxation of
BBN constraints are the
$^4$He and $^7$Li abundance estimates.  Here the systematic uncertainties
are probably largest.  Below I report on our recent analysis of
this situation.

\section{Systematic Uncertainties and $^4$He and $^7$Li}

Recently, several groups have assessed more carefully the
systematic uncertainties present particularly
in the primordial
\he abundance estimates\cite{oliveste,copidns,sass}, and have quoted
various new upper limits on cosmological parameters based on their
 assessments.  It is very clear, based in part on the differing
 estimates, that it is quite difficult at the present time to get an
 accurate handle on these uncertainties.

Because of this, and because we could utilize
the full statistical machinery we previously developed when comparing
predictions to ``observations", we felt it would be useful to prepare
a comprehensive table of constraints on $N_{\nu}$ and
$\Omega_{B}$ for a relatively complete range of different
assumptions about light element abundances.  In so doing, this allowed
us to explore the role of different estimates in the
constraints, as well as the effect of correlations as the light
element abundance estimates vary.  In addition, it allows us to
address several points which we feel are important to
consider when deriving cosmological constraints using BBN
predictions.  Finally, this analysis leads to new
 simple relations between the light element abundances and
limits on cosmological parameters such as the number of neutrinos
, $N_{\nu}$,  and the baryon to photon ratio
$\etaten$.

\subsection{ BBN Predictions and Observations: Systematics,
Correlations and Consistency}

If systematic uncertainties in the inferred primordial
element abundances are dominant, one might wonder whether one need
concern oneself with the proper handling of statistics in the
predicted range.  There is, after all, no well defined way to treat
systematic uncertainties statistically.  For example, should one treat
a parameter range governed by systematic uncertainties as if it were
gaussianly distributed, or uniformly distributed?  The latter
is no doubt a better approximation--i.e. a large deviation within some
range may be as equally likely as a small deviation. But how should one
handle the distribution for extreme values?  Clearly it cannot remain
uniform forever.

Thankfully, there are two factors which make the comparison
of predictions and observations less ambiguous in the case of BBN:

(1) Because the allowed range in the observationally inferred
abundances is much larger than the uncertainty in the predicted
abundances, any constraint one deduces by comparing the two depends
merely on the upper { \it or} lower observational limit for each
individual element, and not only both at the same time.  Thus, one is
not so much interested in the entire distribution of allowed abundances
as one is in one extremum of this distribution.

(2) Systematic uncertainties dominate for the
observations, while statistical uncertainties dominate for the
predictions.

Both of these factors suggest that a conservative but still well
defined approach involves setting {\it strict} upper limits on
$Y_p$, D+$^3$He, and $^7$Li, and a lower limit on D, which incorporate
the widest range of reasonably accepted systematic uncertainties.
 Determining what is reasonable in
this sense is of course where most of the ``art" lies.  Nevertheless,
once such limits are set and treated as strict bounds, then one can
compare correlated predictions with these limits in a well defined way.
In this way one replaces the ambiguity of properly treating the
distribution of observational estimates with what in the worst case may be
a somewhat arbitrary determination of the extreme allowed observational
values.

Clearly all the power, or lack thereof, in this procedure lies in
the judicious choice of observational upper or lower limits.  Because
of our concern about the ability at present to prescribe
such limits I consider a variety of possibilities here.
Nevertheless, once one does choose such a set,
it is inconsistent not to use all of it throughout in deriving
constraints.  If one uses one observational upper limit for
$Y_p$, for example, to derive constraints on the number of neutrinos,
but does not use it when deriving bounds on the baryon density, then
probably one has not chosen a sufficiently conservative bound on $Y_p$
in the former analysis.   It has been argued that a weak,
logarithmic, dependence of
$Y_p$ on
$\etaten$ invalidates its use in deriving bounds on the latter
quantity.  Not
only can this argument be somewhat misrepresentative for an
interesting range of
$Y_p$ values, but until
$Y_p$ exceeds statistically derived upper limits by a large amount, it can
continue to play a signifcant role in bounding $\etaten$ from above.

\subsection{Abundance Estimate Uncertainties: The Range}

(a) $^4$He:  By correlating $^4$He abundances with metallicity for
various heavy elements including O,N and C, in low-metallicity HII
regions one can attempt to derive a ``primordial"
abundance defined as the intercept for zero metallicity.  This can be
determined by a best fit technique, assuming some linear or quadratic
correlation between elemental abundances
\cite{PTP,pagelsimon,pagelterl,walker,oliveste}.
The statistical errors
associated with such fits are now small.  Best fit values obtained
typically range between .228-.232, with statistical ``1$\sigma$" errors on
the order of .003.  This argument yields the upper limit of .24
\cite{walker}
which has been oft quoted in the literature. The key
systematic uncertainty which interferes with this procedure is the
uncertainty in the $^4$He abundance determined for each individual
system, based on uncertainties in modelling HII regions,
ionization, etc used to translate observed line strengths into mass
fractions.  Many observational factors come into play here, and people
have argued that one should add an extra systematic uncertainty of
anywhere from .005-.015 to the above estimate.  Clearly thus, one
should examine implications of
$^4$He abundances in the range .24-.25.  Our work shows that for
$Y_p$ above .25; (a) \he becomes unimportant for bounding $\etaten$,
and (b) the effect on bounds on $N_{\nu}$ can be obtained by
straightforward extrapolation from the data obtained for the range
.24-.25.

(b) $^7$Li:  It is by now generally accepted that the primordial
abundance of $^7$Li is closer to the Spite Pop II plateau than the
Pop I plateau.  Nevertheless, even if one attempts to fit the
primordial abundance by fitting evolutionary models to the Pop II
data points\cite{con}, assuming no depletion,
one still finds a 2$\sigma$ upper limit as large as
 $ 2.3 \times 10^{-10}$.  The role of rotationally induced depletion is
still controversial.  It is clear some such depletion is expected,
and can be allowed for\cite{pinn},
but observations of $^6$Li,
which is more easily depleted, put limits on the amount of
$^7$Li depletion which can be allowed.  We assumed an extreme
factor of 2 depletion as allowable, and thus we
explore how cosmological bounds are affected by a $^7$Li upper
limit as large as $ \approx 5 \times 10^{-10}$.

(c) D and D+$^3$He:  We took the solar system D abundance
of
$ 2
\times 10^{-5}$ as a safe firm lower bound on D, and the previously
quoted upper limit of $10^{-4}$ as a firm upper limit on D+$^3$He
\cite{Yang}.   The results which occur if the D limits are
revised to account for the recent Songaila {\em et al} (1994)
 result were described earlier.

\subsection{Results and Analysis}

Tables 2-4 give our key results. Again, the data were obtained using 1000
Monte Carlo BBN runs at each value of
$\etaten$.  In each case the
number of runs which resulted in abundances which satisfied the joint
constraints obtained by using combinations of the upper limits on
$^4$He, $^7$Li, and D+$^3$He or the lower limit on D was determined.
Limits on parameters were determined by varying these until less than
50 runs out of 1000 (up to
$\sqrt{N}$ statistical fluctuations) satisfied all of the
constraints.

\bc {Table 2: \he Abundance Estimates \& $N_{\nu}$ limits}
\bt{||c||c|lcccr||}  \hline
$Y_p$& $N_{\nu}^{max}$ &\mc{5}{c||}{\#
allowed models:} \\&&
\mc{5}{c||}{\{$^4$He \& [D+$^3$He]\}($^4$He:D+$^3$He)}
\\
 \hline &&&$\eta_{10}$=2.75&2.80&2.85&2.90  & \\
.240
&3.07&&40(603:148)&52(429:254)&46(268:376)&38(170:534)& \\
 \hline &&&$\eta_{10}$=2.80&2.85&2.90&2.95  & \\
 .241
&3.14&&38(532:171)&46(354:309)&39(219:470)&35(131:625)& \\
\hline
.242
&3.21&&41(562:154)&55(451:276)&53(272:423)&52(163:616) & \\
 \hline
.243
&3.29&&17(588:110)&32(410:220)&46(266:378)&36(184:513) & \\
\hline
.244
&3.36&&30(669:102)&44(501:187)&38(353:336)&40(216:464) & \\
\hline
&&&$\eta_{10}$=2.85&2.90&2.95&3.00  & \\
.245&3.43&&50(598:173)&68(449:296)&64(308:427)&54(173:586) & \\
\hline
.247
&3.59&&27(635:84)&30(480:184)&47(338:306)&39(185:488) & \\
 \hline
&&&$\eta_{10}$=2.95&3.00&3.05&3.10  & \\
.250
&3.82&&45(491:207)&47(364:374)&50(225:495)&32(131:587) & \\
 \hline
\et \ec

\bc {Table 3: \he and $^7$Li Abundance Estimates \& $\eta_{10}$
limits}

\bt{||l||c|c||c|c||}  \hline
$Y_p^{max}$&$\eta_{10}^{max}$& \# allowed
models: &
$\eta_{10}^{max}$ &\# allowed
models:
\\
&($^7$Li$_{-10}<$
2.3)&\{$^4$He \& $^7$Li\} ($^4$He:$^7$Li)&($^7$Li$_{-10}<$ 5)&
\{$^4$He \& $^7$Li\} ($^4$He:$^7$Li) \\
\hline .240&3.26&56 (60:998)& 3.26&56 (60:1000)  \\
\hline .241&3.55&45 (45:986)& 3.55&45 (45:1000)  \\ \hline
.242&3.89&45 (47:905)& 3.89&47 (47:1000)  \\ \hline
.243&4.26&50 (60:626)& 4.27&46 (46:1000) \\ \hline
.244&4.64&48 (92:296)& 4.71&49 (49:1000) \\ \hline
.245&5.01&45 (211:118)&5.23&62 (62:984) \\ \hline
.246&5.23&51 (679:62)&5.80&46 (50:810) \\ \hline
.247&5.25&52 (997:52)&6.36&48 (80:500) \\ \hline
\et \ec

\bc {Table 4: $^4$He, D and $^7$Li Estimates \& $\eta_{10}$
limits ($^7$Li$_{-10}<$5;
$D_{-5}>2$)}

\bt{||l||c|c||}  \hline
$Y_p^{max}$&$\eta_{10}^{max}$& \# allowed
models:
\\
&&\{$^4$He \& D \& $^7$Li\} ($^4$He \& D:$^4$He \&
$^7$Li:D \& $^7$Li)
($^4$He:D:$^7$Li)\\
\hline
.248&6.94&48 (136:53:156) (178:516:203) \\ \hline
.249&7.22&52 (177:101:64) (654:217:136) \\ \hline
.250&7.24&47 (191:113:47) (995:191:113) \\ \hline
\et \ec
\bigskip

Table 2 displays the upper limit on  $N_{\nu}$ for various values of
$Y_p$.  As is shown, this is governed by the combination of $^4$He
and D+$^3$He upper limits.  Shown in the table are the number of
acceptable runs out of 1000 when the two elemental bounds are
considered separately and together, for an $\etaten$ range which was
 found to maximize the number of acceptable models. Throughout the
$Y_p^{max}$ region from .24 to .25, both the $Y_p$ and D+$^3$He
limits play a roughly equal role in determining the maximum value of
$N_{\nu}$.  We are able to find a remarkably good analytical fit for
the maximum value of $N_{\nu}$ as a function of $Y_p$ as follows:
\begin{equation}
N_{\nu}^{max} =3.07 + 74.07(Y_p^{max} - .240)
\end{equation}
The linearity of this relation is striking over the whole region
from .24 to .25 in spite of the interplay between the two different
limits in determining the constraint.
Note also that this relation differs from that quoted in
Walker {\em et al}
 between $Y_p$ and $N_{\nu}$ in that the slope we find
is about
$13\%$ less steep than that quoted there.  The two formulae are not
strictly equivalent in that the one presented in
Walker {\em et al}
presented the best fit value of $Y_p$ determined in terms of
$N_{\nu}$, while the present formula gives a relation between the
maximum allowed values of these parameters, based on limits
on the { \it combination} $Y_p$ and D+$^3$He, and on the width of
the predicted distribution.  In this sense, eq. (1) is the
appropriate relation to utilize when relating bounds on $Y_p$ to
bounds on $N_{\nu}$.

Tables 3 and 4, which display the upper bounds on $\etaten$, are
perhaps even more enlightening.  They demonstrate the sensitivity of
the upper limit on
$\etaten$ and hence
$\Omega_{B}$ to the various other elemental upper limits
as
$Y_p$ is varied.  Several features of the data are striking.  First,
note that $^4$He completely dominates in the determination of the
upper limit on
$\etaten$ until
$Y_p$ =.245, even for the most stringent chosen upper limit on
$^7$Li.  If this limit on $^7$Li is relaxed, then $^4$He dominates
as long as the upper limit on $Y_p \le$.248!  Also note
that the ``turn on" in significance of the $^7$Li contribution to
the constraint is somewhat more gradual than the ``turn off" of the
$^4$He constraint.  The former turns on over a range of $\etaten$ of
about 2, while the latter turns off over a range of about 1-1.5.
This gives one some idea of the size of the error introduced in
determining upper bounds by using either element alone, rather
than the combination.  Next, for a
$Y_p$ upper limit which exceeds .248, the lower bound on D begins to
become important.  It quickly turns on in significance so that by the
time the upper limit on $Y_p$ is increased to .25, $^4$He essentially
no longer plays a role in bounding $\etaten$.  Finally, note that both
the relaxed bound on
$^7$Li and the D bound converge in significance at about
the same time, so that for $\etaten > 7.25$, both constraints are
significantly violated.   This implies a ``safe" upper limit on
$\etaten$ at this level, which corresponds to an upper bound
$\Omega_{baryon} \le .163$, assuming a Hubble constant in excess of
40 km/sec/Mpc.   We
again stress that a value this large is only allowed if
$Y_p$ exceeds .250.   If, for example, $Y_p \le .245$, then the
upper bound on $\Omega_{B}$ is essentially completely determined
by
$^4$He and is then at most 0.11.  These limits may be compared to
recent estimates of $\Omega_{baryon}$ based on
X-ray determinations of the baryon fraction in clusters\cite{whiteetal}.

One final comment on the role of $Y_p$ in constraining $\etaten$:
It has been stressed that because of the logarithmic dependence of the
former on the latter, $Y_p$ cannot be effectively used to give a
reliable upper bound on $\etaten$.   While the logarithmic dependence
is correct, any logarithmic relation becomes linear over a
sufficiently small range, and the question of importance then
becomes, how small is ``sufficiently small" in this case.  Using both the
unrelaxed and relaxed upper limit on $^7$Li, the relationship between
the maximum values of $\etaten$ and $Y_p$ remains fairly linear out to
$Y_p^*$ =.245.  Even out to $Y_p$ as large as .248, where
the D and relaxed $^7$Li bounds begin to take over, a quadratic fit
remains good to better than 5$\%$.  The
best linear fit (up to $Y_p^*$) is given by
\begin{equation}
\eta_{10}^{max} \approx 3.22 + 354(Y_p^{max} - .240)
\end{equation}
Seen in these terms, the $\etaten$ upper limit is approximately 4
times more sensitive to the precise upper limit chosen for $Y_p$
than is the $N_{\nu}$ upper limit.  Thus, while there is no doubt that
varying the upper limit on $Y_p$ has a more dramatic effect on the
upper bound one might derive for $\etaten$ than it does for
constraining $N_{\nu}$,the quantitative nature of the
relative sensitivities is perhaps better displayed, for the relevant
range of $Y_p$, by the two relations derived here than it is by
merely saying that one dependence is exponential and the other is
linear.   More important, even recognizing the increased sensitivity of
$\etaten$ on
$Y_p$, unless one is willing to accept the possibility of a rigid upper
bound on $Y_p$ greater than .247, it is overly conservative to ignore
$^4$He when deriving BBN bounds on $\etaten$.

\section{Conclusions:}
BBN has the potential of providing among the strongest
existing constraints on various types of dark matter.  It already
provides the most compelling evidence against the existence of a closure
density of baryonic dark matter.  Whether it is compatible with even a
galactic halo density of baryons depends crucially on our ability to
infer primordial light element abundances with less uncertainty than
presently exists.  As has been
shown, if uncertainties on $^4$He, D, and $^7$Li can be reduced, BBN could
essentially fix the baryonic abundance in the universe, and constrain a
great deal of other ``dark matter" physics beyond the standard model.
Nevertheless, the fact that systematic uncertainties dominate at present
does not block our ability to make statistically meaningful
statements.
Also, as time proceeds and more independent observations are made we
will undoubtedly get a better handle on the
uncertainties which presently limit the efficacy of BBN
constraints.  Until then, the tables and relations presented here
should allow individuals to translate their own
limits on the
light element abundances into meaningful bounds on $N_{\nu}$ and
$\etaten$.

\bigskip
I would like to thank Peter Kernan for
his many essential contributions to the results presented here,
and also David
Schramm and Terry Walker for useful comments about various issues.
Finally, I want to thank the organizers of the UCLA meeting for making it
so exciting and pleasant.


\vskip 0.1in
\noindent{\bf References}

\begin{thebibliography}{99}
\bibitem{kk1}
P. Kernan and L.M.Krauss, Phys. Rev. Lett. {\bf 72}, 3309 (1994)
\bibitem{kk2}
L.M.Krauss and P. Kernan, Ap. J., in press (1994)
\bibitem{kk3}
L.M.Krauss and P. Kernan, Case Preprint, CWRU-P9-94, submitted to Ap. J.
\bibitem{who}
R.V. Wagoner, W.A. Fowler and F. Hoyle, Ap. J. {\bf 148}, 3 (1967);
H. Reeves et al, Ap. J. {\bf 179}, 909 (1973)
\bibitem{kawano}
L.Kawano, Fermilab-Pub-88/34-A (1988); 92/04-A (1992)
\bibitem{helium}
Part. Data Group, {\em Review of Particle Properties}, Phys.Rev.{\bf
D45} (1992); Phys. Rev. {\bf D50} (1994) 1173
\bibitem{walker}
T.P.Walker {\it et al}, Ap.J., {\bf 376}, 51 (1991)
\bibitem{con}
C.P. Deliyannis, P. Demarque, S.D. Kawaler, L.M. Krauss and  P. Romanelli,
Phys Rev. Lett. {\bf 62}, 1583 (1989)
\bibitem{kraussrom}
L.M.Krauss and P.Romanelli, Ap.J., {\bf 358}, 47 (1990)
\bibitem{smithetal}
M.K.Smith, L.H.Kawano and R.A.Malaney, Ap.J.Supp. {\bf 85}, 219 (1993)
\bibitem{gai}
T.Motobayashi {\em et al}, Rikkyo RUP 94-2 / Yale-40609-1141
\bibitem{dic:accbbn}
D.A.Dicus, E.W.Kolb, A.M.Gleeson, E.C.G.Sudarshan,
V.L.Teplitz and M.S.Turner, Phys.Rev., {\bf D26}, 2694 (1982)
\bibitem{kernan}
P.Kernan, Ph.D. thesis, The Ohio State University (1993).
\bibitem{KSW}
P.Kernan, G.Steigman and T.P.Walker, in preparation.
\bibitem{seck:accbbn}
D.Seckel, Bartol Preprint BA-93-16 hep-ph/9305311, Phys. Rev. D, in press
\bibitem{turndod}
S.Dodelson, and M.Turner, Phys.Rev. {\bf D46}, 3372 (1992)
\bibitem{ftd:accbbn}
B.D.Fields, S.Dodelson, and M.Turner, Phys.Rev., {\bf D47},4309 (1993)
\bibitem{Kolbturn}  E. W. Kolb and M. S. Turner
{\em The Early Universe}, Addison-Wesley (Reading, MA. 1990)
\bibitem{turner}
E.W. Kolb , M.S. Turner, A. Chakravorty, and  D.N. Schramm, Phys. Rev. Lett.
{\bf 67} 533 (1991)
\bibitem{kkkssw}
 M. Kawasaki, P. Kernan, H-S. Kang, R.J. Scherrer, G.Steigman
and T.P. Walker, OSU-TA-5-93 (1993)
\bibitem{Walker}
This issue was also
briefly raised by T.P. Walker, Proc. Texas Symp. 1992, C. Akerlof and  M.
Srednicki eds (Ann. N.Y. Acad Sci, 1993)
\bibitem{fullermal} G.M. Fuller, R. A. Malaney, Phys. Rev. {\bf D43} 3136
(1991)
\bibitem{krauss}
L.M.Krauss, Phys.Lett.B {\bf 263}, 441 (1991)
\bibitem{kraternap}
L.M.Krauss, J.Terning and T.Applequist, Phys.Rev.Lett. {\bf 71}, 823
(1993)
\bibitem{dodelmal} S. Dodelson, and L.M. Widrow, Phys Rev. Lett. {\bf 72},
17 (1994)
\bibitem{fuller}
 K. Jedamzik {\it et al}, ASTROPH-9312066 (1994)
\bibitem{olive91}
Olive, K.A., Steigman, G. and Walker, T.P,
{\em Ap. J.} {\bf 380}, L1 (1991)
\bibitem{pagel92}
Pagel, B.E.J. and Kazlauskis, A. {\em MNRAS}, {\bf 256} 49P(1992)
\bibitem{pagel93}
Pagel, B.E.J. {\em Proc. Nat. Acad. Sci.}, {\bf 90} 4789 (1993)
\bibitem{SCHR}
A.Songaila, L.L.Cowie, C.J.Hogan and M.Rugers, Nature {\bf 368}, 599
(1994)
\bibitem{dear}
Dearborn, D.S.P., Schramm, D.N. and Steigman, G.
{\em Ap.J.}, {\bf 302}, 35 (1986)
\bibitem{olive90}
Olive, K.A. { \it et al.}
{\em Phys. Lett.} {\bf B236}, 454 (1990)
\bibitem{oliveste}
Olive, K.A. and Steigman, G. {\em preprint} UMN-TH-1230/94 (1994)
\bibitem{copidns}
T.Copi, D.N.Schramm, M.S. Turner {\em preprint} U. Chicago (1994)
\bibitem{pagelsimon}
Pagel, B.E.J., Simonson, E.A., Terlevich, R.J., and Kennicutt Jr,
R.C., {\em Mon. Not. R. Astr. Soc.} {\bf 255}, 325 (1992)
\bibitem{pagelterl}
Pagel, B.E.J. Terlevich, R.J. and Melnick, J. {\em P.A.S.P.} {\bf 98},
1005 (1986)
\bibitem{PTP}
Peimbert,M and Torres-Peimbert, S. {\em Ap J} {\bf 193}, 327 (1974)
\bibitem{pinn}
Pinsonneault, M.H. {\it et al}
{\em Ap. J. Supp} {\bf 407} 699 (1992)
\bibitem{sass}
Sasselov, D.D. and  Goldwirth, D. {\em preprint} astro-ph 9407019 (1994)
\bibitem{Yang}
Yang, J. {\it at al}  {\em Ap.J.} {\bf 281}, 493 (1984)
\bibitem{whiteetal}
White, S.D.M.{\it et al.}{\em
Nature} {\bf 366},429 (1993)
\end{thebibliography}
\end{document}